# Machine Learning Coupled Trajectory and Communication Design for UAV-Facilitated Wireless Networks

Aksh Garg

**Abstract** Augmenting wireless networks with Unmanned Aerial Vehicles (UAVs), commonly referred to as drones, offers a promising avenue for providing reliable, cost-effective, and on-demand wireless services to desired areas. However, existing UAV communication and trajectory schemes are inefficient as they assume limited drone mobility and static transmission power. Furthermore, they tend to rely upon convex approximations to highly non-linear functions and fail to adopt a combination of heuristic and convex methods. This paper considers a Multi-UAV system where UAV-mounted mobile base stations serve users on the ground. An iterative approach using block gradient descent is used to jointly optimize user scheduling, UAV trajectories, and transmission power for maximizing throughput over all users. Subsequently, an innovative technique for initial trajectory predictions was developed using a K-means clustering algorithm for partitioning users into subgroups and a genetic algorithm for initializing shortest flight paths within clusters. Finally, convex optimization solvers such as MATLAB's Fmincon are used for fine-tuning parameters. Extensive simulation and optimization results demonstrate a 33.57%, 87.4%, and 53.2% increase in system throughput for the 1, 2, and 3 UAV scenarios respectively when compared to existing trajectory and communication design schemes. Furthermore, the K-means and genetic algorithm reveal additional improvements in throughput by around 15%. Our results note diminished increases in throughput for increases in UAV trajectory period as the period approaches higher values. Further research into joint adoption of convex and non-convex schemes as well as consideration of environment-dependent channel models would allow for a faster and more optimal deployment of UAVs.

*Index Terms*—5G, Genetic Algorithm, Heuristic Algorithms, K-Means Clustering, Optimization, Unmanned Aerial Vehicles, Wireless Communications

## I. INTRODUCTION

THE rapid increases in user data requests and consumption have resulted in strains on existing wireless communication frameworks. Future 5G networks are expected to support 1000-fold greater traffic loads, 100 billion connected devices, and new requirements on latency, connectivity, reliability, etc. [1] Effectively visualizing such an environment, however, would require a rapid development in current technological methods and techniques.

Unmanned aerial vehicles (UAVs), commonly referred to as drones, are expected to be an important component of the upcoming wireless networks that can potentially help bridge this communication gap. Compared to conventional communication systems with fixed architectures, characteristics of UAVs such as their freedom of movement, flexibility in deployment, and ability to establish line-of-sight communication links make them ideal for administering a versatile means of cellular coverage [2]. For instance, in contrast to terrestrial base stations, wherein the deployment is primarily two dimensional, UAVs have the ability to adjust their altitude, allowing for greater likelihood of establishing line of sight (LoS) communication links to ground users (Fig 1) Furthermore, UAV base stations can supplement existing communication frameworks by providing on-demand, wireless ad-hoc services during special events or decreasing network saturation in hotspot areas. Other important applications of UAVs include their ability to facilitate Internet of Thing (IoT) Scenarios, whose devices have short coverage and weak links. Several companies have already begun capitalizing upon this opportunity and are investing in projects, such as Aquila by Facebook and Project Loon by Google.

Aksh Garg is a student at Palos Verdes Peninsula High School, 27118 Silver Spur Rd, Rolling Hills Estates, CA 90274, USA. (e-mail: akshgarg@gmail.com)

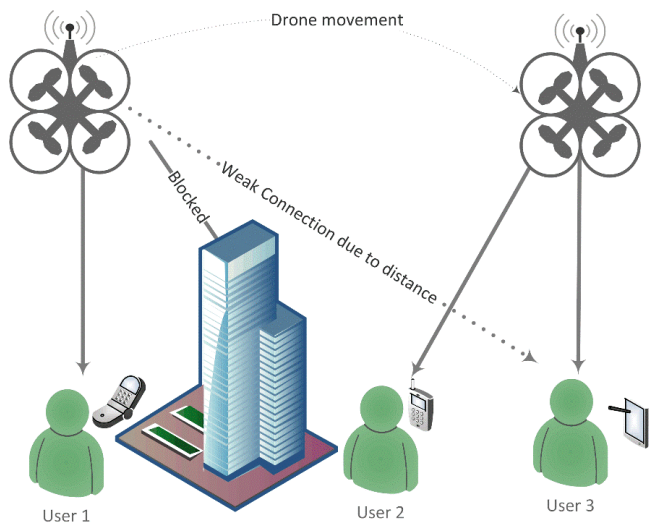

Figure 1: Demonstration of Line of Sight issues in wireless networks. The signal is unable to reach users who don't have a direct line of sight as the signal is too weak to penetrate through boundaries. Moreover, the signal strength fades as the distance between the transmitter (UAVs) and receivers (users) increases. UAVs help mitigate such challenges by virtue of their mobility and ability to move closer to users they intend to serve and establish line-of-sight links. Such characteristics are especially important as we move towards 5G communication with mm-waves as they have extremely short travel distances.

Similarly, extensive undertakings have been taken in academia to determine the optimal means of deploying UAVs for the goal of enhancing currently existing cellular networks. In particular, the static deployment problem of UAVs for wireless networks has been well studied. The authors in [3] present an analytical technique for optimizing the altitude of Low-altitude aerial platforms (LAPs) for maximizing radio coverage on the ground. They show that the optimal altitude is a function of statistical parameters and allowed loss in the environment. In [4], the authors expand the work in [3] to the case of two-interfering UAVs. They study the impact of distance between two UAVs on the coverage area and analytically derive the optimal distance between the two drones for ensuring maximum coverage. The work in [5] derives the downlink coverage probability of UAVs as a function of altitude and antenna gain. They subsequently determine the minimum number of UAVs needed to guarantee certain coverage probabilities. The authors in [6] explore a UAV-augmented HetNet, which is more adaptive to changes in spatio-temporal traffic distributions. They investigate the problem of optimal 3D deployment of UAVs for maximizing the number of served ground users. The authors in [7] attempt to find the minimum number of drone-BSs and their placement to ensure all users are served using a heuristic algorithm. The authors in [8] also attempt to solve the problem of minimizing the number of UAVs needed provide wireless coverage in a region; however, they do so using a polynomial-time algorithm, which greatly accelerates algorithm convergence time. In [9], the authors highlight the properties of the drone-cell placement problem and formulate it as a 3-D placement problem with the objective of maximizing the revenue of the network. The work in [10] focuses on the deployment of UAVs for energy-efficient communication and data collection. They explore the optimal 3D deployment of UAVs, device-UAV associations, and power control with the goal of minimizing total uplink transmit power.

Although static UAV placement is an important problem, in order to realize the full potential of UAV-enabled wireless networks, one must capitalize upon the UAV's mobility. Although trajectory optimization for UAVs has usually been explored from a mechanics/control perspective [11-15], recent work has directed attention towards the interplay between trajectory optimization and UAV's role in wireless communications. Such a problem, however, is complex to solve as it relies on many physical constraints and parameters. For instance, when designing UAV trajectories, factors such as channel variations due to mobility, limited battery life, and flight constraints must be accounted for [2]. Furthermore, analytically solving for UAV trajectories is challenging as it involves finding solutions to an infinite number of optimization variables. For instance, the work in [16] focused on finding the optimal trajectory for a single UAV system equipped with multiple antennas for maximizing mean-rate in uplink communications. In [17], the authors iteratively optimize UAV transmit power and trajectories for maximizing throughput within the system. The authors in [18] minimize the total UAV energy consumption while covering the user area by computing an optimal set of waypoints and optimal speed of UAVs between such waypoints. Similarly, in [19], considering collision avoidance, fuel efficiency, and communication criteria, optimal UAV paths are found using mixed-integer linear programming. In [20], UAV location and movement are optimized to improve connectivity in Ad-Hoc (temporary and on-demand) network provisioning services assuming that the UAVs have complete information about the user device location. The authors in [21] derive a theoretical model on the propulsion energy consumption of fixed-wing UAVs as a function of the UAV's flying speed, direction and acceleration. They, then, introduce an efficient design for maximizing energy efficiency with general constraints on the trajectory. [22] considers the case of a UAV flying cyclically over the ground terminals at a fixed altitude. The time allocations are optimized to maximize their minimum throughput while simultaneously minimizing user-access delay. Their results reveal a fundamental tradeoff between throughput and access delay in the cyclical multiple access scheme. The work in [23] optimizers user scheduling, UAV trajectories and transmission power for maximizing the minimum average throughput in the system.

Although the works [16], [17], [18], and [22] provide seminal work in the field of trajectory optimization, they are limited in their consideration of a single-UAV system. Although a single UAV system has clearly defined advantages, it has limited capability as a result of its practical size, weight and power constraints [24]. This calls for a more comprehensive model with multiple UAVs for a more efficient assimilation of UAVs into wireless network schemes. Furthermore, while [19], [20], [21] consider the multi-UAV case, they fail to consider proper strategies for the reduction of interference from additional UAVs by appropriate UAV transmit power control. Such a consideration is especially fundamental in maximizing the signal to noise ratios (SINRs) for ground users, which in turn sets an upper limit upon the channel capacity by the Shannon Hartley Theorem [25]. For example, without adjusting UAV transmit power, there are high levels of interference between UAVs in close proximity to each other, which in turn decreases the quality of information transferred and overall throughput. With an adjustable power, such interference may be decreased, thus, allowing for shorter and less distant trajectories.

In this paper I consider a multi-UAV system where UAV-mounted mobile base stations serve users on the ground, with the goal of maximizing the mean-rate over all users. The UAVs are assumed to share the same frequencies and therefore interference management between the drones is thoroughly considered both when designing and evaluating the solution. Moreover, the UAVs are assumed to be equipped with omnidirectional Antennas, which further amplifies the importance of interference management within the system. A machine learning coupled joint optimization of user scheduling, UAV trajectory, and transmit power algorithm is proposed to attain the goal of maximizing throughput to ground users. Particularly, a combination of heuristic algorithms such as the genetic algorithm and machine learning algorithms such as K-means clustering are used as a preliminary step to solving the formulated non-convex optimization problem. The solutions obtained through these algorithms are then passed as input arguments to a nonlinear optimization solver such as Fmincon's interior point algorithm. Such a combination of heuristic and machine learning techniques with a nonlinear optimization setup has not been previously investigated in literature to my best knowledge.

The proposed setup has significant advantages over other models that either fail to jointly optimize the three variable blocks or rely solely upon convex optimization schemes for converging to solutions. We can note the former by considering the case of a simple trajectory optimization without power transmit control. An optimal UAV trajectory design is able to ensure powerful, short-distance, LoS dominated links between UAV-user pairs, thereby increasing throughput. However, in cases where the UAVs need to serve users that are densely populated in a particular region on the ground, a trajectory optimization setup in itself may only account for high interference among UAVs by increasing the inter-UAV distance. This, however, also results in lower received signal strength and corresponds to lower system throughput. Joint trajectory and transmit power optimization, on the other hand, allows two UAVs to be in close proximity to each other by reducing the signal transmit power of one of them. In a case where transmit power is not optimized, the solution obtained from a simple trajectory optimization problem serves as the lower bound. Therefore, joint trajectory and transmit power optimization is guaranteed to result in greater throughput than simple trajectory optimization schemes.

The importance of the K-means clustering and the genetic algorithm is reflected when designing the initial UAV trajectories. Examination of solutions found using existing trajectory optimization methods reveals that users have tendencies of being partitioned into distinct clusters that are then more effectively served by UAVs. The heuristic methods are thus designed to take advantage of this UAV-User clustering. Doing so, greatly accelerates the solution procedure as the heuristic algorithms for initializing trajectories have much lower computational complexity than gradient-based optimization as done in conventional approaches. Moreover, consider the close coupling between user scheduling, UAV trajectories, and transmit power. To make the problem more tractable to solve, an iterative approach is taken where each block is alternatively and sequentially optimized. Therefore, one block is optimized while holding the other two constants. Although this approach greatly reduces problem complexity and is able to converge to favorable solutions, the problem framework is still non-convex in nature, which raises the concern of the algorithm converging to non-optimal local minima. The addition of the K-means and genetic algorithm improve the initial guesses to Fmincon and thus increase the chances of converging to minima with higher throughputs. Overall, the heuristic algorithms play fundamental roles in the expediting solution procedure and increasing overall throughput. Although, not guaranteeing convergence to a global minimum, such a technique therefore allows us to find a time-efficient sub-optimal solution with acceptable throughputs.

The rest of this paper is organized as follows; Section II introduces the system model and problem formulation. Section III introduces and explains the proposed algorithmic design and optimization procedure. Section IV presents and discusses the results. Finally, section V summarizes key points and concludes the paper.

## II. SYSTEM MODEL

### A. System Model

*The system model presented below is adopted from [23]. The model is presented again for convenience. The work in this paper differs in its solution procedure. Particularly, it attacks the problem in its non-convex state instead of finding convex approximations to highly nonlinear and nonconvex function. It adopts methods of K-means clustering and genetic algorithm for improving initial guesses for solutions and addresses the problem in its non-convex form without using sequential convex optimization techniques to convert to a convex form.*

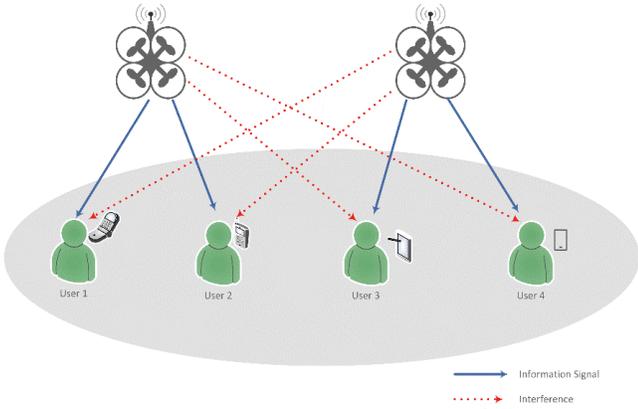

Figure 2: A multi-UAV enabled wireless Network. We observe that each user receives signals from multiple drones, some of which may be unrequested and thus act as noise. The blue lines in the figure represent signals from UAVs that are serving a particular user, whereas the red lines represent unwanted signals coming from UAVs that aren't directly serving a user due to the omnidirectionality of their antennas.

I consider a system where $M \geq 1$ UAVs (drones) are used for providing service to a set of $K \geq 1$ stationary users on the ground. The set of UAVs is denoted $\mathcal{M}$ and the set of users denoted $\mathcal{K}$. $|\mathcal{M}| = M$ and $|\mathcal{K}| = K$, where $|X|$ denotes the cardinality of set X. Furthermore, all UAVs are assumed to be identical and therefore, each UAV can serve any user in the system. The UAVs are also assumed to share the same frequency band for communication which makes interference management an important problem to simultaneously consider when designing UAV trajectories, user schedules, and transmission powers.

The UAVs are assumed to cyclically serve all users with a period of T. This means that the UAVs must return to their starting point at the end of each cycle. Moreover, the choice for the period plays an important role in the system performance. On one hand, a larger period T allows the UAVs to move closer to the users it is serving and establish better channels for communication, whereas, a shorter period prevents UAVs from moving closer to desired users and corresponds to lower throughput gains. On the other hand, larger periods of T also correspond to greater user-access delays as each user has to wait longer until it is served by the UAV with a strong communication channel again. This tradeoff between throughput and user-access delay is noted again in the results section of the paper, wherein we provide results with different values for the period to note its impact upon the performance of our system.

A 3D cartesian model of the system is adopted, wherein $w_k = \{x_k, y_k\} \in \mathbb{R}^{1 \times 2}$, denotes the horizontal coordinates of each user $k$. All users are assumed to be located at a fixed altitude, which I define to be 0 for convenience. The UAVs are also assumed to fly as a fixed height $H$. Such an assumption keeps the problem complexity in manageable levels, although variable UAV altitude may be considered in future work, albeit at a much higher computational cost. The horizontal components of the UAVs position at time $t$ are denoted by $q_m(t) = \{x_m, y_m\} \in \mathbb{R}^{1 \times 2}$, for $0 \leq t \leq T$. Since the UAV trajectories were assumed to be periodic, the location of the UAV $m$ at time T must equal its starting location, i.e,

$$q_m(at\ initial\ time) = q_m(T), \forall\ m. \quad (1)$$

In practice, the trajectories are limited by physical constraints on the UAV such as its maximum speed, $V_{max}$. For collision avoidance purposes, a minimum distance $d_{min}$ must always be maintained between UAVs. This gives us another set of constraints:

$$\|\dot{q}(t)\| \leq V_{max}\ \forall\ 0 \leq t \leq T, \quad (2)$$
$$\|q_m(t) - q_j(t)\| \geq d_{min}\ \forall\ m, j\ \epsilon\ M\ s.t.\ m \neq j. \quad (3)$$

The period is discretized into N equal sub slots indexed by n = 1, 2, …, N. An elemental time slot $\Delta t$ thus equals $\frac{T}{N}$. Making this substitution, the UAV trajectory may be represented by a set of N 2-dimensional sequences $q_m[n] = [x_m[n], y_m[n]]$, and the problem formulated as follows:

$$q_m[1] = q_m[N],\ \forall\ m,j, m \neq j, \quad (4)$$
$$\|q_m[n] - q_j[n]\| \geq d_{min},\ \forall\ m,j, m \neq j, and \quad (5)$$
$$\|q_m[n+1] - q_m[n]\| \leq V_{max}\Delta t\ \forall\ n = 1, \dots, N-1. \quad (6)$$

The distance from user k to UAV $m$ is given by
$$d_{k,m}[n] = \sqrt{H^2 + \|q_m[n] - w_k\|^2}. \quad (7)$$

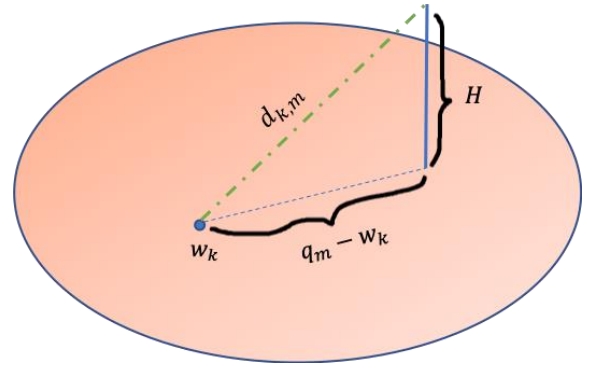

FIGURE 3: UAV-USER DISTANCE REPRESENTATION. FIGURE CREATED BY AUTHOR

Furthermore, the case where communication links are dominated by LoS (Line of Sight Links) is assumed. In this case, the channel quality may be approximated using a free space path loss model, for which the channel gain only depends upon the distance between the transmitter and the receiver. In such as case, the channel power gain from UAV $m$ to user k may be expressed by $h_{k,m}[n]$ using the Friis free space equation, which says

$$\frac{P_r}{P_t} = G_t G_r * \frac{\lambda^2}{(4\pi d)^2} \Rightarrow \frac{P_t}{P_r} = \frac{G_t G_r \lambda^2}{(4\pi)^2} * \frac{1}{d^2},$$

where $G_t$ represents the transmitter antenna directivity, and $G_r$ represents the receiver antenna directivity.

Making the substitution $\frac{G_t G_r \lambda^2}{(4\pi)^2} = \rho_o$ and $\frac{P_t}{P_r} = h$, we obtain

$$h_{k,m}[n] = \frac{\rho_0}{H^2+\|q_m[n]-w_k\|^2}, \quad (8)$$

Where $\rho_o$ represents the channel power gain at a reference distance of 1m.

I now define a binary variable $\alpha_{k,m}[n]$ representing whether or not the user k is being served by UAV *m* during a time slot n. Particularly, a user k is served by UAV *m* if $\alpha_{k,m}[n] = 1$, and $\alpha_{k,m} = 0$ if a user k is note being served by UAV *m*.

I assume that in a given unit of time, each UAV is only capable of communicating with a single ground user and that each user is being served by at most 1 UAV. Therefore,

$$\sum_{k=1}^{K} \alpha_{k,m}[n] \leq 1 \ \forall \ m,n, \quad (9)$$
$$\sum_{m=1}^{M} \alpha_{k,m}[n] \leq 1 \ \forall k,n, \quad (10)$$
$$\alpha_{k,m}[n] \in \{0,1\} \ \forall \ k,m,n. \quad (11)$$

The signal throughput over a given channel is limited by the Shannon limit of wireless communications which states that the max rate of information transfer over a given frequency is given by $\log_2(1 + \gamma_{k,m}[n])$, where $\gamma_{k,m}[n]$ represents the signal to noise ratio (SINR) at user k. To calculate the SINR for a user k, we introduce another set of variables $p_m[n] \leq P_{max}$ representing the downlink transmission power of UAV *m* at a given time slot n. If a user k is being served by the UAV, i.e. $\alpha_{k,m}[n] = 1$, the corresponding signal to noise ratio may be represented by

$$\gamma_{k,m}[n] = \frac{h_{k,m}[n]p_m[n]}{\sum_{j=1, j \neq m}^{M} h_{k,j}[n]p_j[n] + \sigma^2}, \quad (12)$$
$$p_m[n] \leq P_{max} \ \forall \ n,m. \quad (13)$$

Here $h_{k,m}$ represents the channel power gain from UAV *m* to user k as found in equation (8). The numerator represents the received signal from the desired UAV, whereas the denominator represents the co-channel interference experienced due to the transmissions of other UAVs at time n.

### B. Problem Formulation

Let $A = \{\alpha_{k,m}[n], \forall k,m,n\}$ represent the user schedules for all users $k \in K$ and UAVs $m \in M$ from n = 1, ..., N. Similarly let $Q = \{q_m[n], \forall m,n\}$ and $P = \{p_m[n], \forall m,n\}$. We assume that the fixed positions of ground users $W = \{w_k, \forall k\}$ are known and the goal is to maximize mean rate among all users by jointly optimizing user scheduling $A$, UAV trajectories $(Q)$, and transmit power $(P)$ over all time slots. This maximized mean rate is referred to as max-mean rate through the rest of the paper. The problem may thus be expressed as:-

$$\max_{A,Q,P} \frac{1}{N} \sum_{k=1}^{K} \sum_{n=1}^{N} \sum_{m=1}^{M} \alpha_{k,m}[n] \log_2(1 + \gamma_{k,m}[n]) \quad (14)$$
$$subject \ to \ constraints \ (4),(5),(6),(9),(10),(11),(13)$$

Solving this optimization problem is difficult because 1) the optimization variables A are binary and thus involve integer constraints, and 2) the variables A, Q, and P are closely coupled and difficult to simultaneously optimize.

## III. PROPOSED ALGORITHM

To make the problem more tractable to solve, we first relax the binary variable $\alpha$ into continuous variables. Relaxing into a continuous state allows us to use analytic gradients more easily. Dealing with such gradients is a lot easier with continuous variables rather than discrete ones. Moreover, our optimization solver, Fmincon is designed to work with continuous functions with continuous first derivatives, which makes it important to relax variables to a continuous form. Further we address the joint optimization problem by solving for each variable block while holding the other two constant. Specifically, User Scheduling A may be optimized for a given UAV trajectory Q and Transmit Power P. Similarly, for a given user scheduling setup A and transmit power P, the trajectory Q may be optimized, and the same for P. In order to accelerate the solution search process, I employ a K-means algorithm for determining the optimal clustering of users for each UAV and then a genetic algorithm for minimizing UAV travel distance in initial paths within the clusters.

### A. Trajectory Guess

For a given communication system with a set of $M$ UAVs and K ground users, the users are partitioned into $M$ distinct clusters using a K-means clustering algorithm. This allows the users within a cluster to be more efficiently and distinctly served by an individual UAV. Moreover, by virtue of the K-mean clustering algorithm's nature, not only is the distance between the individual users and the centroid of each clutter minimized, but the centroids themselves are also pushed farther away from one another. This further reduces possible chances of interference due to larger distances between UAVs in different clusters. The Figure below represents the K-means algorithm running on a scenario with 50 users and 3 UAVs. As can be noted in the Fig, the algorithm effectively splits the users into distinct groups/clusters.

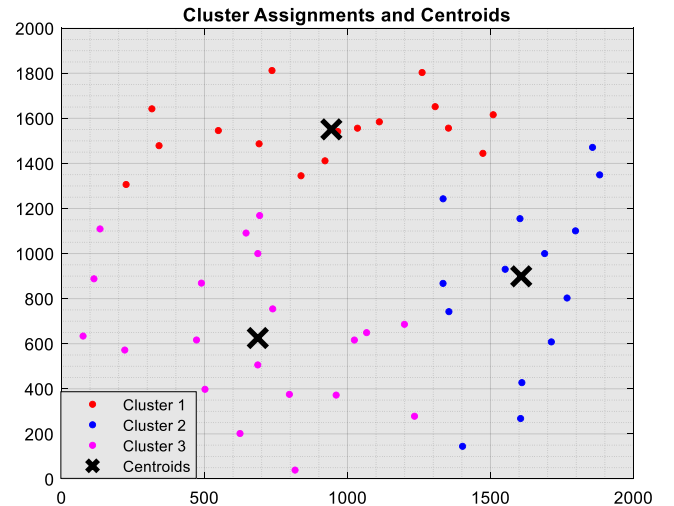

Fig 4: Demonstration of K-Means Clustering Algorithm. The algorithm successfully partitions the users into distinct clusters that would then be served by a single UAV. Properties of K-Means, such as minimizing the distance of each user from the centroid allows UAVs to have shorter trajectories, which, in turn, improves channel quality as UAVs can move closer to their users. Furthermore, the greater distance between centroids decreases the interference among UAVs in different clusters.

Post clustering, the initial paths for the UAVs must still be initialized. In order to generate our initial trajectories, we attempt to find the shortest path the UAVs can take while visiting each user at least once. Doing so, however, is difficult and bears similarities to the Travelling Salesman Problem, which involves identifying the shortest path that a salesman may take if he must travel to a predetermined set of cities, such that the overall distance that he travels is minimized. This problem, however, is known to be NP-hard in general, meaning that the entire solution space must be exhaustively searched before reaching a solution. Several methods have been proposed to identify the optimal means of minimizing this travel distance including greedy search, memetic algorithms, and simulated annealing, among others. These algorithms, however, tend to converge to suboptimal solutions or suffer from extremely high computational times. To increases chances of reaching a more optimal solution and expediting the solution process, I employ a genetic algorithm for finding the minimum travelling distance for the UAVs within a given cluster.

**FIG 5.A: INITIALIZED TRAJECTORIES WITHOUT GENETIC ALGORITHM**

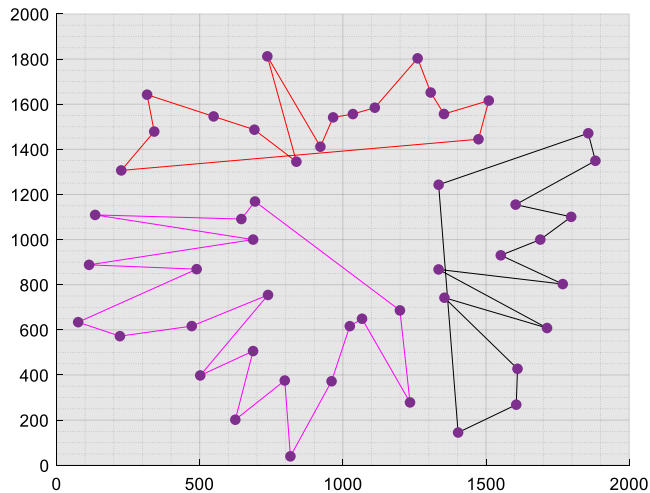

**FIG 5.B: Initialized Trajectories with Genetic algorithm**

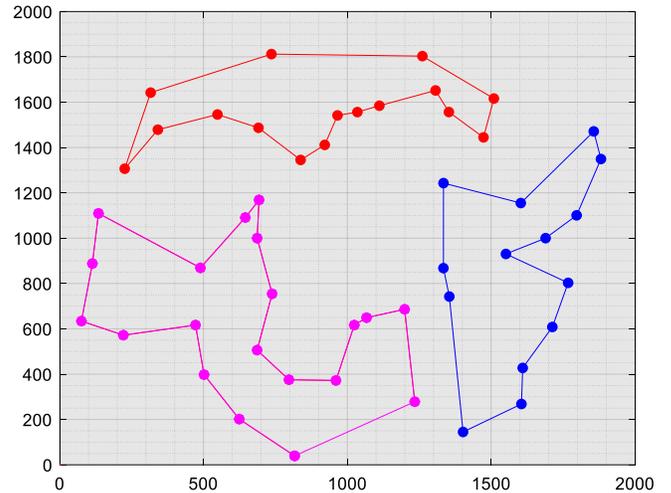

Fig 5: Comparison of Initialized trajectories with and without a genetic algorithm. We note that the trajectories found without using a genetic algorithm (Fig 5.a) are significantly longer and contain several overlapping paths. In contrast, the trajectory generated using the genetic algorithm (Fig 5.b) is significantly shorter and avoids overlap between various regions. This, in turn, allows the UAVs to spend more time over individual users, increasing overall system throughput. Moreover, shorter flight paths mean that the optimization solvers have greater room to move within the constraints and improve initial guesses. such alterations are more difficult with suboptimal case seen in figure 5.a.

The genetic algorithm described in the preceding sentence, works similar to how genes function in evolution. Particularly, each gene corresponds to a guess towards a solution for the problem we are trying to solve, which in this case corresponds to minimizing the travel distance for the UAVs (or commute distance for the salesman). Next, several of these genes are generated to form a set known as a population (similar to how several individuals form a population in biological concepts of evolution). The objective function (total travel distance) is evaluated for each of these genes and the genes corresponding to more optimal solutions (those having higher throughputs) have higher chances of being transmitted onto future generations. As genes are transmitted onto future generations, changes are introduced to some genes in the form of random mutations (changes in the vectors representing the solutions) or crossover events (exchanges between two parent genes that are used to create new children genes with attributes from both parents). These changes add more variety to the gene pool and allow the algorithm to explore diverse regions of the solution space. Over time, the population selects better and better genes until it converges to a stop depending on iteration limit or relative gains in solution quality.

Additional conditions must be checked post the genetic algorithm. Particularly, the total distance travelled by the UAV in the current guess might not always be feasible. Since UAV speeds are limited, the UAV is only capable of travelling a maximum distance of $V_{max} \cdot T$ in a time interval T. In the case where the total distance calculated is greater than the maximum distance allowed, the trajectory is scaled down, maintaining its current center until the distance is permitted under the system requirements.

The combination of heuristic and traditional optimization methods represents the most fundamental and core distinction from traditional optimization schemes. Although most schemes attempt to find convex approximations to highly nonlinear and nonconvex functions, doing so excludes the possibility for every converging to the true optimal minimum. The adopted solution procedure in this paper targets the problem in its true form without making any approximation of sorts. However, this invokes the risk of converging to poor local minimum. By making solutions found through traditional optimization schemes, one may note the users in the scenario tend to be split into clusters that are then served by individual UAVs. Taking advantage of such characteristics, the K-means and genetic algorithms effectively split the users in the scenario into groups. This expedites the solution process greatly as the heuristic algorithms have much lower computational complexity than traditional optimization frameworks, while simultaneously increasing the chances of reaching high throughput solutions. In fact, the proposed framework can provide high throughput solutions as presented in the results section even as the number of time-steps is reduced to 150-300.

### B. User Scheduling and Association

For a given set of UAV trajectories Q and Transmit Power P, we may optimize the user schedules and associations through the following setup:

$$\max_A \frac{1}{N} \Sigma_{k=1}^K \sum_{n=1}^N \sum_{m=1}^M \alpha_{k,m}[n] \log_2(1 + \gamma_{k,m}[n]), \quad (15)$$
$$subject\ to\ constraints\ in\ equations\ (9),(10),(11)$$

This formulated problem is a well-known Linear Programming problem and may be solved used standard optimization solvers such as MATLAB's Fmincon.

### C. Trajectory Optimization

For a given set of User Schedules A and Transmit Power P, the UAV trajectories may be solved by solving the following optimization problem:

$$\max_Q \frac{1}{N} \Sigma_{k=1}^K \sum_{n=1}^N \sum_{m=1}^M \alpha_{k,m}[n] \log_2(1 + \gamma_{k,m}[n]) \quad (16)$$
$$subject\ to\ constraints\ (4),(5),(6)$$

This given formulation is highly non-convex in nature. For such a problem, the system consists of multiple feasible regions and local optima. Consequently, for such a problem formulation, the initial trajectory guesses play a major role in determining accurate problem convergence. A general analysis of existing optimized trajectories reveals that trajectories tend to take a closed loop path passing through all users in the cluster. Therefore, the proposed initial guess in PART I serves as an efficient means of approximating the initial trajectories. Given a decent initial guess, the optimization algorithm presented above can be solved using common optimization solvers such as MATLAB's Fmincon. Furthermore, the proposed algorithm may be simply extended to other UAV use cases by exchanging the objective function in (14) with the goal of the new problem one is trying to solve.

### D. Transmit Power Control

With a given user association and scheduling, $A$, and UAV trajectories, $Q$, the UAV power transmission may be optimized by solving

$$\max_P \frac{1}{N} \Sigma_{k=1}^K \sum_{n=1}^N \sum_{m=1}^M \alpha_{k,m}[n] \log_2(1 + \gamma_{k,m}[n]) \quad (17)$$
$$subject\ to\ constraint\ (13),$$
$$where\ \gamma_{k,m}[n] = \frac{h_{k,m}[n]p_m[n]}{\sum_{j=1,j\neq m}^M h_{k,j}[n]p_j[n] + \sigma^2}.$$

The given problem is also non-convex in nature but may be solved using gradient based approaches such as MATLAB's Interior Point algorithm.

### E. Overall Algorithm Design

I use a block gradient-descent approach as used in [23] for optimization purposes, wherein the variable blocks are individually and iteratively optimized. The sequential optimization is continued until the gains obtained through each individual iteration is lower than a certain threshold limit, which may be adjusted depending upon the accuracy of the solution desired. The overall algorithm is additionally presented in a simplified manner below:

```
Initialize UAV trajectories Q{1}, using
a K-means based approach.
Initialize A{1} and P{1} either randomly
or using the methods presented in the
paper.
Iteratively optimize each individual
variable block:
    For i = 1,2,…, Number Of Desired
Iterations:
        Solve problem (15) over {A{i},
    Q{i}, P{i}} to obtain A{i+1}
        Solve problem (16) over {A{i+1},
    Q{i}, P{i}} to obtain Q{i+1}
        Solve problem (17) over {A{i+1},
    Q{i+1}, P{i}} to obtain P{i+1}
```

## IV. RESULTS

This section presents the numerical results to demonstrate the effectiveness of the proposed algorithm. Basing the environmental parameters on commonly used statistical variables from [17], [21], [22], [23] and conventional UAV properties such as those outlined in [26], I consider a system where users are randomly distributed over a 2x2 km² area. The UAVs in the simulation are assumed to have a max velocity of 50 m/s and a max transmission power of 0.100W. We assume a free space noise loss of -110dBm or $10^{-14}$ W, and a channel gain of -60dB ($10^{-6}$) at 1m. A minimum inter-UAV distance of 50m is considered necessary for collision avoidance. The follow results are obtained for two major scenario types: those with 6 and 10 users respectively. Most of the results analyze the case of 6 UAVs due to computational limitations of a personal laptop computer. We conclude this section with example of the algorithm running on varying user distributions to show its effectiveness.

### A. Single UAV Case

I first consider the scenario of a single UAV, i.e. $M = 1$, where there is no inter-UAV interference. In this scenario, the optimization problem could be simplified to a special case of algorithm 1, where the UAV transmit power is kept fixed at its maximum, i.e. $p_m[n] = P_{max} \forall n$. Thus, the given problem reduces to a simpler case of problem (14), where A and Q need to be optimized while P remains constant. We first note the effect of T on the UAV trajectory. I observe that as T gets larger, the UAVs get more and more time to get closer to and establish better channels of communication with their users. They do so by quickly speeding to the user they intend to serve and hovering over that user for a period of time, so as to allow better communication channels and faster communication rates. This, however, comes at the expense of a greater user access delay, for each user is made to wait a longer period of time before being served strongly by a UAV again. For instance, when T is large (210s), the UAVs spend a large amount of time right above each user, and the UAV path simply becomes a linear closed path between the various users. In contrast, in the case of short flight times, T = 60s, the UAVs are unable to remain stationary over individual users and manage to serve all the users in a given area. Therefore, the UAVs continue travelling at near max speed throughout the serving period, so as to uniformly serve all the users in a region. Fig 6 demonstrates the differences between UAV trajectories with a large and small period T respectively. We can note in the figure the dependence upon the system throughput and UAV trajectory time periods. Particularly, we see that in the case of 60s UAV trajectories, a throughput of only 0.962 bps/Hz is attained while in the case of 210s trajectories, throughputs of 1.476 bps/Hz may be obtained. Also made apparent in the Fig is the inability of the UAV to effectively cover the entire serving area in 60s even when flying at maximum speed throughout its trajectory. In contrast, the ability of the UAV to effectively reach each individual user in the 210s case allows for better communication channels and consequently much greater throughputs. Furthermore, we note that at lower values for T, the UAV throughput is more responsive to increases in trajectory period. Particularly, increasing the UAV period to 90s in the 1 UAV case results in the throughput rising to 1.217 bps/Hz, a 0.255 bps/Hz increase in throughput for only a 30s increase in trajectory period. However, When increasing the trajectory period from 90s to 210s (a 120s increase), the throughput only changes by 0.259 bps/Hz. The average rate of change (AROC) for the first step was 0.0085bps/Hz/s (0.255/30) while for the latter the AROC is 0.0022 (0.259/120). Given the marginal increases in throughput for increasing values of period, we would therefore like to direct more attention and resources towards systems that are functioning on extremely short trajectory periods as such systems have the most potential gain.

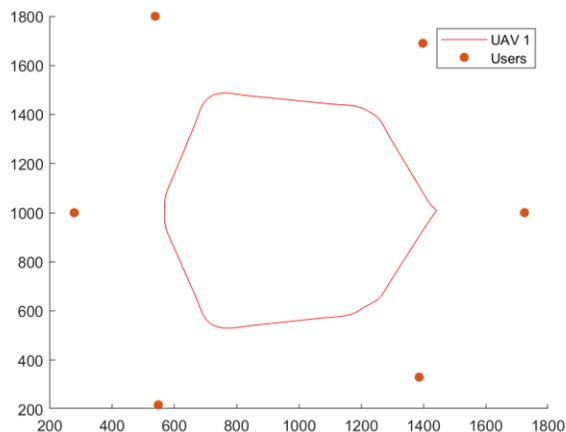

**Fig 6(a): 1 UAV – 6 Users – 60s**
**Max-Mean Rate: 0.962 bps/Hz**

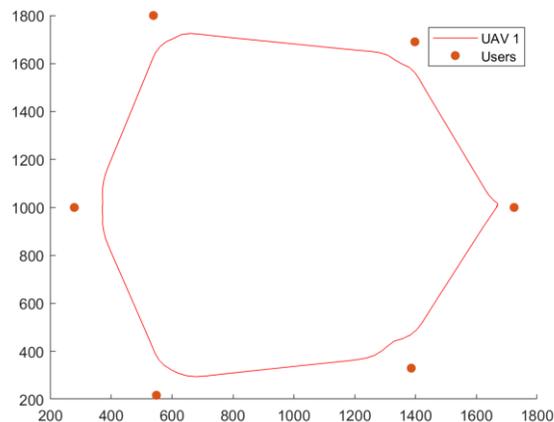

**Fig 6(b): 1 UAV – 6 Users -90s**
**Max-Mean Rate: 1.2173 bps/Hz**

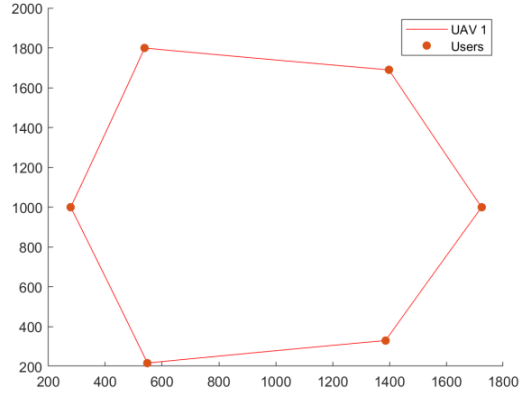

**Fig 6(c): 1 UAV – 6 Users -210s**
**Max-Mean Rate: 1.476 bps/Hz**

Figure 6: The effect of T upon UAV Trajectories. Large values of T result in greater throughput at the expense of greater user-access delay. Fig 6(a) corresponds to T = 60s, fig 6(b) corresponds to T = 90s, and fig 6(c) corresponds to T=210s. We note the changes in UAV trajectories as we move from 60 to 90 to 210s. With shorter trajectory periods the UAV paths are shorter and flight radii smaller. As the period approaches greater values, the UAVs are able to reach individual users, thereby establishing better communication channels and consequently, greater throughputs.

We evaluate the effectiveness of the proposed algorithm under two schemes: Scheme-1: without K-Means clustering initialization, and scheme-2: with K-Means clustering initialization. These two schemes are compared against two standard problem settings: 1) static UAV deployment, and 2) circular trajectories. For each of the scenarios listed above user scheduling was optimized using a given trajectory. Therefore, the only difference in optimization methods for these schemes is in regards to what type of trajectory design approaches were taken as user scheduling remains controlled for. We may also note the important case of static UAV deployment as it is analogous to fixed deployment as done using base stations right now. Consequently, static deployment serves as perhaps the most fundamental benchmark for communication designs. For static deployment, the initial UAV deployment points were initialized by a K-means clustering algorithm and achieved a maximum-transmission rate of 1.0538 bps/Hz. It was further noticed that static UAV deployment was independent of time, i.e. the chosen value for T had no noticeable impact upon system performance. This is expected as the UAV serves all users the same way throughout the period scenario and thus has no reliance upon time due to lack of mobility. The flight paths for circular trajectories were found using circle-packing theory [27] using methods similar to those outlined in [23]. As expected, the circular trajectories did show a dependence upon time, as it influenced how long each UAV could spend near each user as well as the radius of the circles it was able to fly in. We note that in the 1-UAV case, the circular UAV trajectories only marginally outperforms statically deployed UAVs, achieving a max-mean rate of 1.105 bps/Hz. The trajectories found by my proposed algorithm in schemes 1 and 2 show significant increases in throughput for the system. Particularly, scheme 1 manages to increase the max-mean rate to 1.3801 bps/Hz and scheme-2 attains a max-mean rate of 1.476 bps/Hz. Although both schemes show significant increases in throughput in comparison to the static and circular case, scheme 2 outperforms 1 by nearly 0.1 bps/Hz and demonstrates the importance of initial guesses in trajectory optimization.

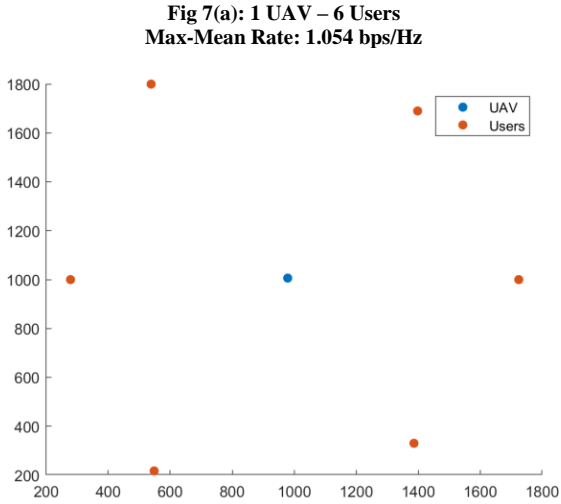
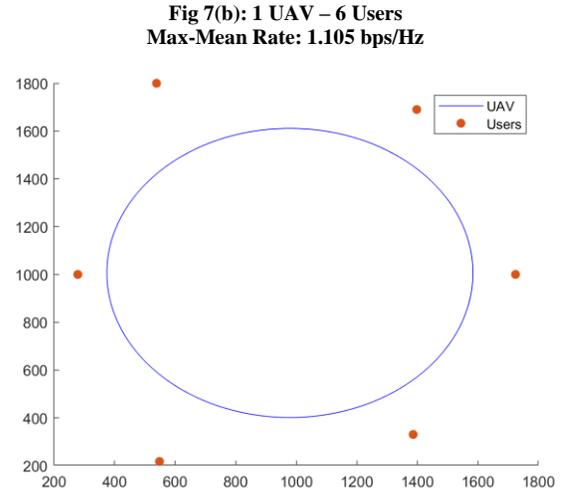
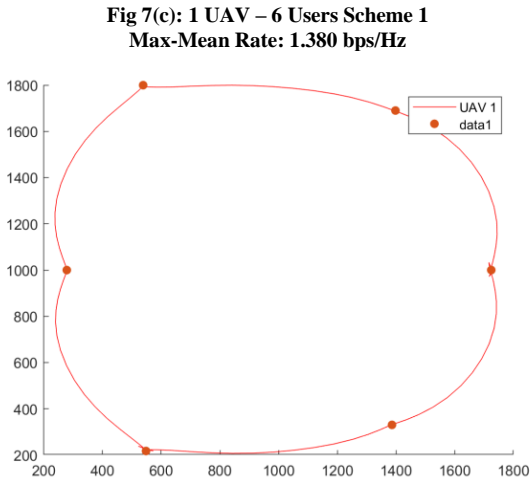
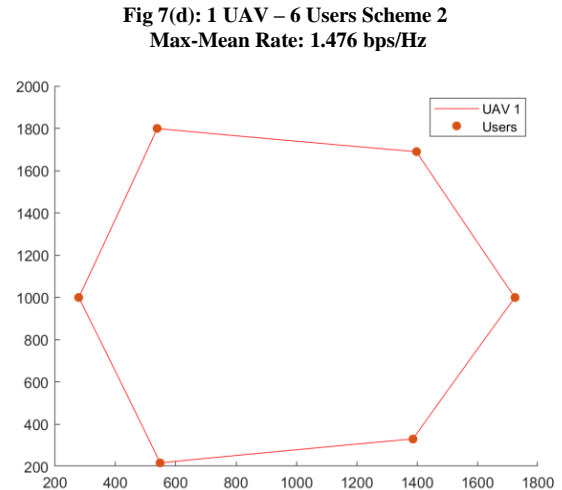

Figure 7: Comparison of various schemes for a 1 UAV wireless network. We note the differences in throughput for the given designs: the Static UAV Scenario (Fig 7(a)) and the circular trajectory design (Fig 7(b)) achieve 1.0538 bps/Hz and 1.105 bps/Hz respectively. The solutions found by my algorithm without (Fig 7(c)) and with (Fig 7(d)) k-means clustering are 1.3801 bps/Hz and 1.476 bps/Hz respectively.

These results demonstrate the effectiveness of my proposed algorithm for optimizing UAV trajectories and user schedules for a single UAV scenario. We note increases in transmission rates from 1.054 bps/Hz in the static UAV case to 1.476 bps/Hz in scheme 2. Therefore, in comparison to current network deployment setups with fixed (static) base stations, optimized UAV deployment as mobile base stations could increase system transmission rate by 40.06%. Furthermore, in comparison to common UAV trajectory planning schemes with circular trajectories, the proposed algorithm shows an increase in throughput by 33.57%. Although I only consider the case of 6 ground users here due to computational limitations, the same algorithm may be used for solving higher complexity problems involving a larger number of users.

### B. Two Interfering UAVs

I now consider the case of 2 UAVs serving a set of 6 users on the ground. In order to take co-channel interference in this scenario into account, UAV transmit power is optimized in tandem with UAV trajectories and user scheduling to ensure best performance. Each UAV is assigned a set of users by partitioning the ground users into the most similar and closely spaced groups by the k-means algorithm discussed in section III.A. Furthermore, similar to the case of a 1 UAV system, the scenario time period T plays a major role in the allowed trajectory states and transmissions. However, for a 2 UAV scenario, user access-delay is highly reduced due to the shorter amounts of time each UAV has to travel to serve its set of users. We notice this effect in Fig 5, where even for T=60s, the UAV is able to efficiently serve the users in its set. Hence the problem of user-access delay is greatly reduced by using multiple UAVs as it allows for shorter flight paths and more direct serving of individual users. In fact, we note that even at trajectory periods of 60s, the 2-UAV system still manages to attain a throughput of 1.497 bps/Hz and with a trajectory period of 90s, a throughput of 1.759 bps/Hz is attained. Such throughputs are fairly close to the throughput of 2.083 bps/Hz attained with a 210s trajectory. We also note the larger rate of increases at lower trajectory periods. The transition from 60 to 90s brings about an AROC of 0.0087 bps/Hz/s ($\frac{1.759-1.497}{30}$) whereas the transition from 90s to 210s brings about an AROC of 0.0027 bps/Hz/s ($\frac{2.083-1.759}{120}$). We also note the progression of UAV trajectory' shape as we move from 60 to 90 to 210s. With greater values for the period, the UAV paths become less restrained by speed and time limitations and can therefore take longer and more complicated shapes.

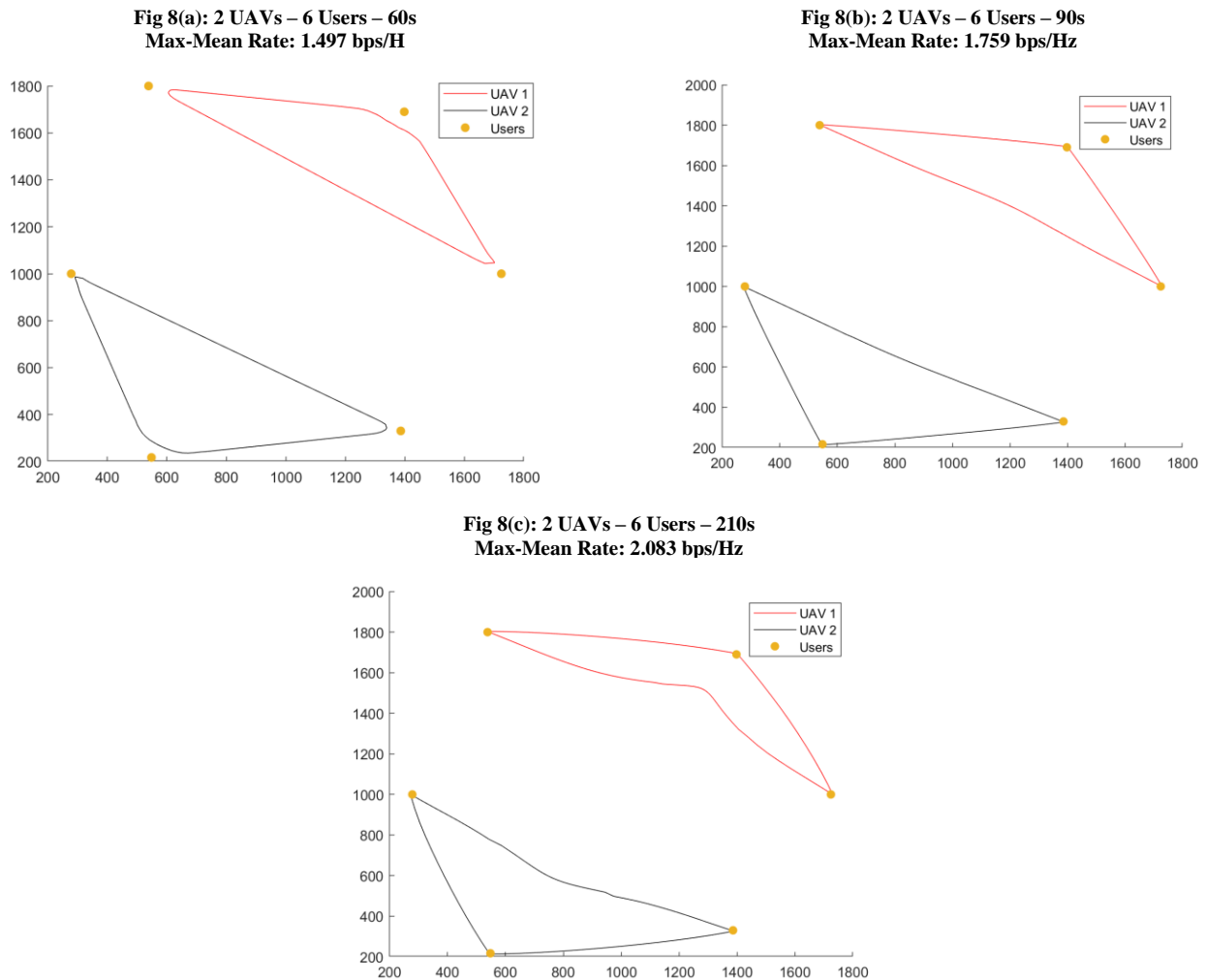

Figure 8: Effect of Time Period (T) on the mean rate. We note that in the case of a 2 UAV scenario, even at a T of 60s (fig 8(a)), the trajectories begin to approach those found under longer durations (Fig 8(c)). This demonstrates the use of multi-UAV networks for not only increasing overall system throughput, but also reducing user-access delay. We additionally note a greater AROC in throughput at lower values for the trajectory, showing that at larger values of t, trajectory time period increases are associated with only marginal increases in throughput.

I now discuss the effectiveness of the proposed algorithm in comparison to various other trajectory design schemes. We consider 4 trajectory design schemes: 1) Static Deployment, 2) Circular Deployment 3) Proposed Algorithm without transmit power optimization 4) Proposed Algorithm with transmit power optimization. The framework for schemes 1 and 2 is the same as discussed in the 1 UAV case above, albeit, with more drones now. In the case of static deployment, we observe a max-mean rate of 1.384 bps/Hz. However, we notice that when trying to maximize the average user rate, the model penalizes certain users. In such a case, we have the issue of some users receiving nearly 0 bps/Hz over the simulation period. To offer more promising results, this time trying to maximize a minimum rate over all users. This gives us a max-min rate of 0.7893 bps/Hz, which although smaller than the max-mean rate found provides a uniform coverage to all users in a setting. Thus, depending upon the requirements of the situation, one may choose between the goal of maximizing the mean rate over all users, which we do in this paper, or ensuring each user receives a minimum signal strength. The latter may be done by optimizing over the min-rate. To address this problem of uniform serving of all users in a region, alternative objective functions are also introduced. These include a weighted objective function where lower values of throughput are heavily penalized. This is enforced using a log-weighted objective function which by its nature penalizes low values more than in awards high values. Moreover, since the log-function is continuous rather than discrete, it is more friendly to the optimization process than a minimum function.

Circular deployment schemes, such as current ones relying upon circle-packing theory [27] obtain a max-mean rate of

1.084 bps/Hz. Further my simulation results demonstrate a weakness ensued through using circle packing as a commonly used UAV trajectory initialization scheme. As circle packing fails to consider the idealized centers for UAV trajectories, the suggested paths fail to deliver as influential of results as found through center searching in k-means. This is in fact explicitly noted when even the found solution in the static case, which is boosted by k-means, surpasses that found through circle-packing. Next, the proposed algorithm without power optimization demonstrates a significant increase in User-UAV throughput, raising the max-mean rate to 2.031 bps/Hz. This corresponds to an increase of 46.7% and 87.4% in comparison to the static and circular cases respectively. Note the smaller increase in the static case may be misleading as the static scenario severely punishes unserved users. In comparison to the solution found through a min-rate optimization in the static scenario, we observe an increase of 157.3% in system throughput. Finally, simulations result in 4) show only a small increase from the max-mean rates found in 3, increasing it from 2.031 bps/Hz to 2.083 bps/Hz. This marginal increase, however, is expected as the 2 UAVs are sufficiently far apart to reduce interference effects. Nonetheless, small differences in the UAV paths are still observable for the two scenarios. Particularly, the trajectories found in 4) tend to be flatter when compared to 3). This is as with transmission power control; the UAV trajectories may be closer while still attaining comparable or higher communication rates.

**Fig 9(a): 2 UAVs – 6 Users – Static Deployment**
**Max-Mean Rate: 1.384 bps/Hz**

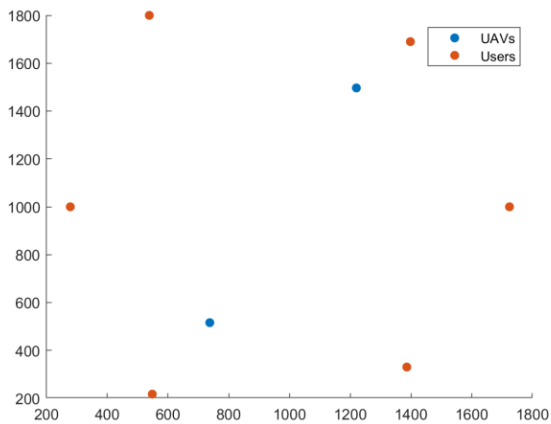

**Fig 9(b): 2 UAVs – 6 Users – Circular Trajectories**
**Max-Mean Rate: 1.084 bps/Hz**

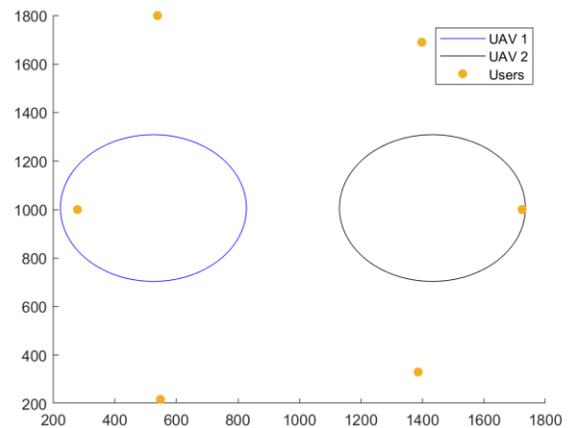

**Fig 9(c)**
**2 UAVs – 6 Users – Proposed Algorithm w/o Power Transmit Control**
**Max-Mean Rate: 2.031 bps/Hz**

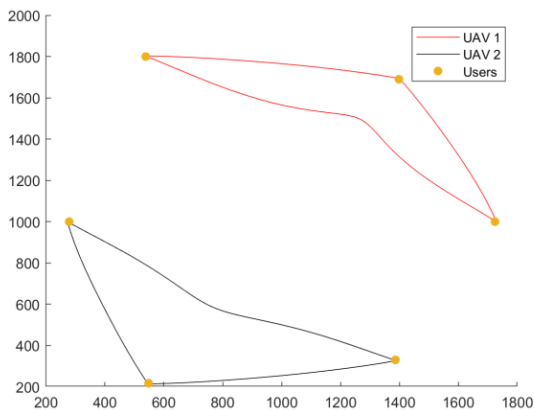

**Fig 9(d)**
**2 UAVs – 6 Users – Proposed Algorithm with Power Transmit Control**
**Max-Mean Rate: 2.083 bps/Hz**

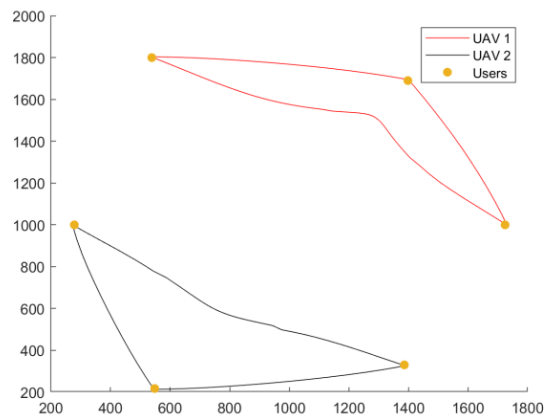

Figure 9: Comparison of various schemes for a 2 UAV wireless network. The static deployment scheme (Fig 9(a)) achieves a throughput of 1.384 bps/Hz. Circular Trajectories (Fig 9(b)) as those found through circle-packing theory achieve throughputs of 1.084 bps/Hz. The trajectories found by my proposed algorithm show marked increases, raising throughput to 2.031 bps/Hz without UAV Transmit Control (Fig 9(c)) and 2.083 bps/Hz with UAV transmit Control (Fig 9(d))

## C. Multi-UAV Analysis

In this section, I consider the case of 3 UAVs serving a set of 6 users on the ground, i.e, $M = 3$. In order to minimize co-channel interference, we employ the join trajectory and power optimization algorithm discussed in section 3. Here we examine and compare the effectiveness of my algorithm to other existing schemes. I consider the following existing schemes: 1) Static Deployment and 2) Circular-trajectories using circle-packing theory and compare them to the trajectory obtained through 3) Joint trajectory and communication design with my proposed algorithm. For each scenario, I optimize user scheduling and UAV transmit power, and therefore the UAV trajectories serve as the primary differentiator among the various methods.

The following results are obtained for the various designs. For the case of static UAV deployment, a max-mean rate of 2.247 bps/Hz is obtained. However, we note the same problem of particular users being severely punished as seen in the 2 UAV scenario, where although the system throughput average is increased, a particular user may be receiving no or a minute amount of signal. The trajectories found through circle-packing theory attain a max-mean rate of 1.712 bps/Hz, which is significantly higher than the throughputs attained through circular trajectories for the 1 and 2 UAV scenario. Finally, capitalizing on the UAVs mobility as done in 3) maximizes the system throughput. Specifically, the algorithm is able to increase the max-mean rate to 2.624 bps/Hz, a 16.8% increase from the static deployment case, and a 53.2% increase from the circular packing design. Further one must note the irregularities suggested by the static case. In reality, it is often infeasible to punish individual users by not providing them sufficient signal amounts. When compared to the solution found through a min-rate optimization scheme (1.112 bps/Hz), we notice an improvement of 136% from the static to mobile case.

Furthermore, like the mobile cases for 1 and 2 UAVs, the UAV trajectory periods play important roles in the throughput of the system for the 3 UAV case as well. However, in the 3 UAV case, the throughput is less strongly lessened at lower trajectory periods. We note that even at $T = 60s$, a throughput of 2.298 bps/Hz is attained. Furthermore, the changes in UAV trajectory shape from 60s to 210s are marginal (with the UAV spending more time hovering over users in the 210s) as the UAV is effectively able to reach the users its attempting to serve even in the 60s case. Therefore, our results show that increasing UAV number is one effective way of minimizing user-access delay in the system.

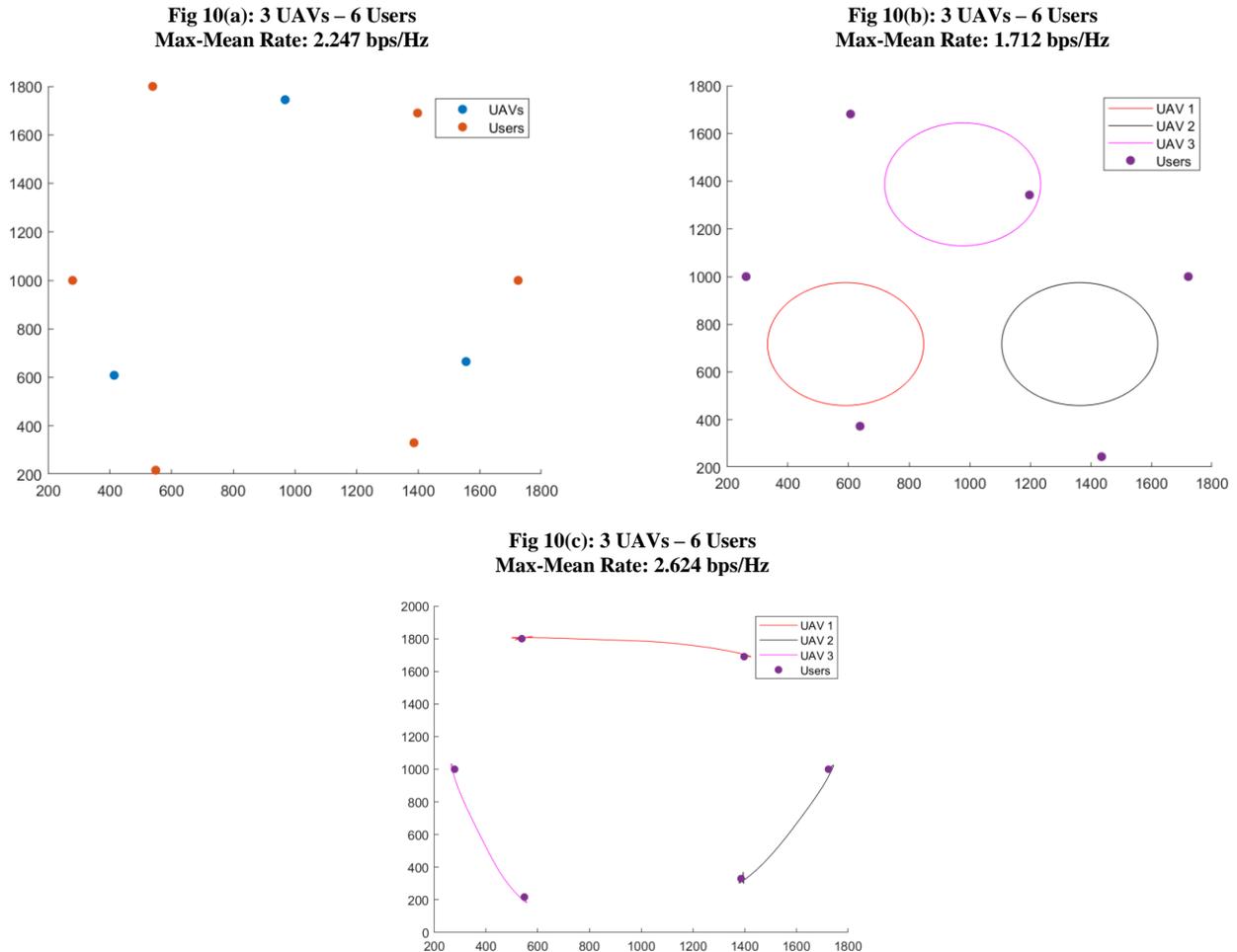

**Fig 10(a): 3 UAVs – 6 Users**
**Max-Mean Rate: 2.247 bps/Hz**

**Fig 10(b): 3 UAVs – 6 Users**
**Max-Mean Rate: 1.712 bps/Hz**

**Fig 10(c): 3 UAVs – 6 Users**
**Max-Mean Rate: 2.624 bps/Hz**

Figure 10: Comparison of trajectory designs for a 3 UAV scenario. The static deployment case(Fig 10(a)) attains a maximum mean-rate of 2.247 bps/Hz. Circular Deployment using Circle-Packing methods (Fig 10(b)) achieves a max-mean rate of 1.712 bps/Hz. The proposed algorithm (Fig 10(c)) surpasses both providing a max-mean rate of 2.624 bps/Hz

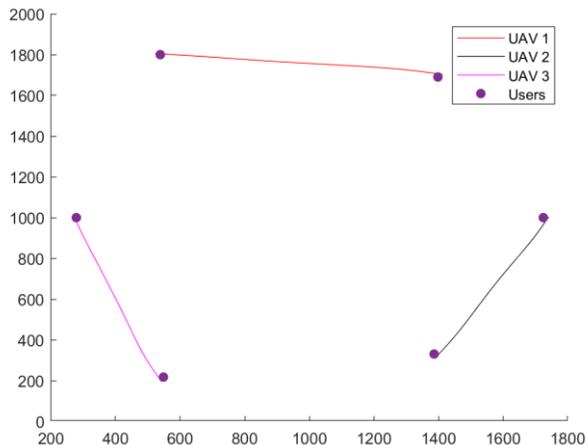

**Fig 11(a): 3 UAVs – 6 Users–60s**
**Max-Mean Rate: 2.298 bps/Hz**

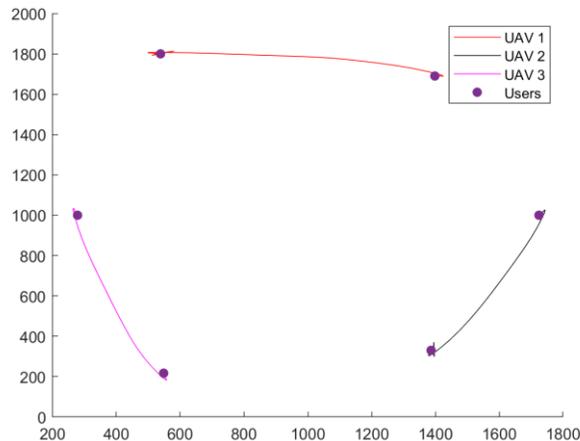

**Fig 11(b): 3 UAVs – 6 Users–210s**
**Max-Mean Rate: 2.624 bps/Hz**

Fig 11: The impact of period on a 2 UAV system. We note that even in the 60s trajectory case (Fig 11(a)), the system throughput is fairly high with a max-mean rate of 2.298 bps/Hz. An increase in UAV period to 210s (Fig 11(b)) allows the trajectory curvature to increase and UAVs spend more time hovering over individual users as represented by the knots in the figure. Moreover, the increase in period by 150s only results in a throughput increase of 0.326 bps/Hz. This shows that a high number of UAVs effectively reduce the UAV period and throughput tradeoff.

*D. Solution Performance under different random User Initializations*

Here I present various alternate user positioning setups where the problem convergence and solution behavior were tested. The results demonstrate the versatile applications for which the proposed framework is applicable. The diverse user distributions help illustrate the capability of the algorithm to perform in variable user distributions. Moreover, as the number of users increases, we can see that the UAV paths become more and more complicated as noted in fig 12(d). Expanding these results to include a greater number of users is another area of exploration to consider.

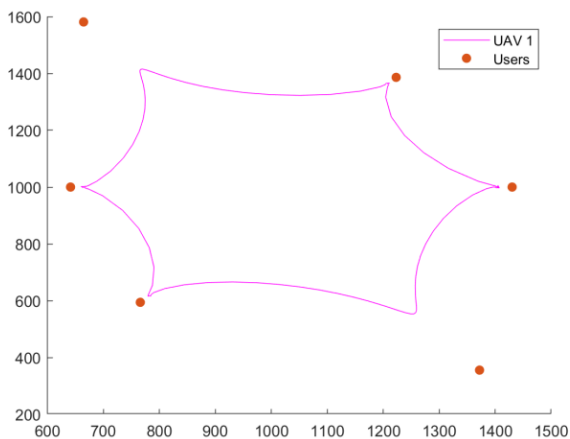

**Fig 12(a): 1 UAV – 6 Users – Traditional Schemes**
**Max-Mean Rate: 0.875 bps/Hz**

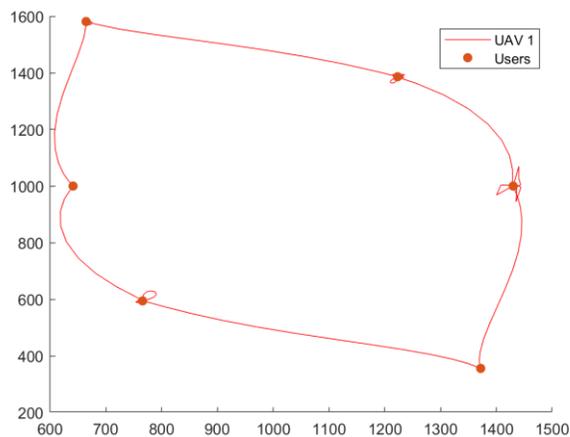

**Fig 12(b): 1 UAV – 6 Users – Proposed Scheme**
**Max-Mean Rate: 1.522 bps/Hz**

**Fig 12(c): 3 UAVs – 6 Users – Proposed Scheme
Max-Mean Rate: 2.563 bps/Hz**

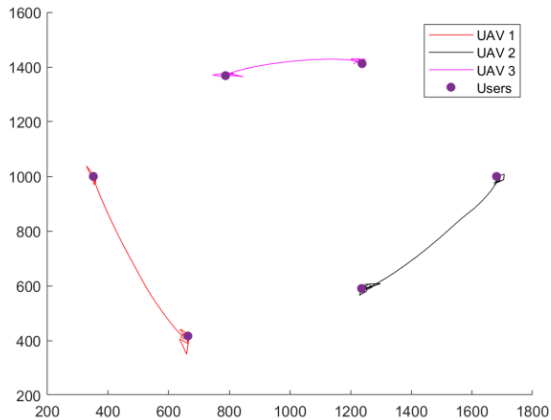

**Fig 12(d): 3 UAVs – 10 Users – Proposed Scheme
Max-Mean Rate: 1.872 bps/Hz**

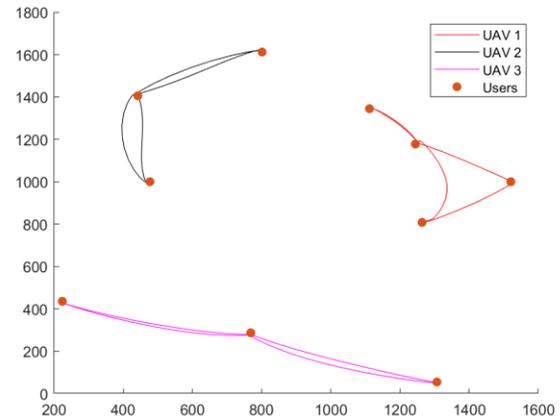

Figure 12: Additional Trajectories found by Proposed Algorithm. Demonstrates the wide variety of scenarios for which the given algorithm is applicable. Fig 12(A) represents a given user scenario solved without proper k-means initialization, and fig 12(b) represents the same scenario with better initial guesses. Fig 12(c) depicts another 6 user random initialization and the throughput that the algorithm provides. Finally, fig 12(d) depicts a 10 user case. once can note the higher detailed paths that the algorithm is able to provide for more complicated scenarios. The max-mean rate is lower in comparison to the 3 UAV and 6 user scenarios as each UAV now has to serve more users on average.

## V. Conclusion

This paper proposed a new method of UAV Trajectory and communication design. The proposed algorithm using both block gradient descent and k-means clustering, and genetic algorithm marks a key distinction to traditional communication design schemes. Whereas traditional approaches tend to find convex approximations to highly nonlinear and nonconvex functions, the approach presented in this paper attacks the problem in its non-convex form, using techniques from Machine Learning (ML) and heuristic designs to improve initial guesses. The algorithm is able to attain significant improvements in throughput gains when compared to other commonly used benchmark schemes. The commonly noticed tradeoff between user-access delay and throughput is observed with greater trajectory periods corresponding to greater user-access delay. Although this work focused its numerical results to the case of 1, 2, and 3 UAVs due to computational limitations, the algorithm may be expanded to solve problems involving greater numbers of UAVs and ground users with greater computational resources or by identifying key traffic density points. Moreover, the problem may be expanded to encompass 3-D flight plans instead of the simpler 2D case considered here. Lastly, this paper considers a simple free-space channel model due to limited computational resources. Considering the effects of alternate channel models including probabilistic line of sign and environmental specific parameters would allow for the development of more robust trajectory designs

## Acknowledgment

This research did not receive any specific grant from funding agencies in the public, commercial, or not-for-profit sectors.

## References

[1] B. Li, Z. Fei and Y. Zhang, "UAV Communications for 5G and Beyond: Recent Advances and Future Trends," in IEEE Internet of Things Journal, vol. 6, no. 2, pp. 2241-2263, April 2019, doi: 10.1109/JIOT.2018.2887086.

[2] M. Mozaffari, W. Saad, M. Bennis, Y. Nam and M. Debbah, "A Tutorial on UAVs for Wireless Networks: Applications, Challenges, and Open Problems," in IEEE Communications Surveys & Tutorials, vol. 21, no. 3, pp. 2334-2360, thirdquarter 2019, doi: 10.1109/COMST.2019.2902862.

[3] A. Al-Hourani, S. Kandeepan and S. Lardner, "Optimal LAP Altitude for Maximum Coverage," in IEEE Wireless Communications Letters, vol. 3, no. 6, pp. 569-572, Dec. 2014, doi: 10.1109/LWC.2014.2342736.

[4] M. Mozaffari, W. Saad, M. Bennis and M. Debbah, "Drone Small Cells in the Clouds: Design, Deployment and Performance Analysis," 2015 IEEE Global Communications Conference (GLOBECOM), San Diego, CA, 2015, pp. 1-6

[5] M. Mozaffari, W. Saad, M. Bennis and M. Debbah, "Efficient Deployment of Multiple Unmanned Aerial Vehicles for Optimal Wireless Coverage," in IEEE Communications Letters, vol. 20, no. 8, pp. 1647-1650, Aug. 2016.

[6] I. Bor-Yaliniz and H. Yanikomeroglu, "The New Frontier in RAN Heterogeneity: Multi-Tier Drone-Cells," in *IEEE Communications Magazine*, vol. 54, no. 11, pp. 48-55, November 2016, doi: 10.1109/MCOM.2016.1600178CM.

[7] F. Jiang and A. L. Swindlehurst, "Optimization of UAV Heading for the Ground-to-Air Uplink," in IEEE Journal on Selected Areas in Communications, vol. 30, no. 5, pp. 993-1005, June 2012.

[8] J. Lyu, Y. Zeng, R. Zhang and T. J. Lim, "Placement Optimization of UAV-Mounted Mobile Base Stations," in IEEE Communications Letters, vol. 21, no. 3, pp. 604-607, March 2017

[9] Yaliniz, E.-K. R. Irem Bor, Amr, and Halim, "Efficient 3-D Placement of an Aerial Base Station in Next Generation Cellular Networks," arXiv.org, 26-Feb-2016. [Online].

[10] M. Mozaffari, W. Saad, M. Bennis and M. Debbah, "Mobile Unmanned Aerial Vehicles (UAVs) for Energy-Efficient Internet of Things Communications," in IEEE Transactions on Wireless Communications, vol. 16, no. 11, pp. 7574-7589, Nov. 2017.

[11] K. Dogancay, "UAV path planning for passive emitter localization," IEEE Transactions on Aerospace and Electronic Systems, vol. 48, no. 2, pp. 1150–1166, 2012.

[12] A. Rucco, A. P. Aguiar, and J. Hauser, "Trajectory optimization for constrained UAVs: A virtual target vehicle approach," in Proc. IEEE

International Conference on Unmanned Aircraft Systems (ICUAS), June 2015.
[13] J. S. Bellingham, M. Tillerson, M. Alighanbari, and J. P. How, "Cooperative path planning for multiple UAVs in dynamic and uncertain environments," in Proc. IEEE Conference on Decision and Control, Dec. 2002.
[14] J. How, Y. Kuwata, and E. King, "Flight demonstrations of cooperative control for UAV teams," in AIAA 3rd" Unmanned Unlimited" Technical Conference, Workshop and Exhibit, 2004, p. 6490
[15] P. Chandler, S. Rasmussen, and M. Pachter, "UAV cooperative path planning," in AIAA Guidance, Navigation, and Control Conference and Exhibit, 2000, p. 4370
[16] F. Jiang and A. L. Swindlehurst, "Optimization of UAV Heading for the Ground-to-Air Uplink," in IEEE Journal on Selected Areas in Communications, vol. 30, no. 5, pp. 993-1005, June 2012.
[17] Y. Zeng, R. Zhang and T. J. Lim, "Throughput Maximization for UAV-Enabled Mobile Relaying Systems," in IEEE Transactions on Communications, vol. 64, no. 12, pp. 4983-4996, Dec. 2016.
[18] C. D. Franco and G. Buttazzo, "Energy-Aware Coverage Path Planning of UAVs," 2015 IEEE International Conference on Autonomous Robot Systems and Competitions, Vila Real, 2015, pp. 111-117, doi: 10.1109/ICARSC.2015.17.
[19] Grøtli, E.I., Johansen, T.A. Path Planning for UAVs Under Communication Constraints Using SPLAT! and MILP. J Intell Robot Syst 65, 265–282 (2012) doi:10.1007/s10846-011-9619-8
[20] Z. Han, A. L. Swindlehurst and K. J. R. Liu, "Optimization of MANET connectivity via smart deployment/movement of unmanned air vehicles," in IEEE Transactions on Vehicular Technology, vol. 58, no. 7, pp. 3533-3546, Sept. 2009.
[21] Y. Zeng and R. Zhang, "Energy-Efficient UAV Communication With Trajectory Optimization," in IEEE Transactions on Wireless Communications, vol. 16, no. 6, pp. 3747-3760, June 2017.
[22] J. Lyu, Y. Zeng and R. Zhang, "Cyclical Multiple Access in UAV-Aided Communications: A Throughput-Delay Tradeoff," in IEEE Wireless Communications Letters, vol. 5, no. 6, pp. 600-603, Dec. 2016.
[23] Q. Wu, Y. Zeng and R. Zhang, "Joint Trajectory and Communication Design for Multi-UAV Enabled Wireless Networks," in IEEE Transactions on Wireless Communications, vol. 17, no. 3, pp. 2109-2121, March 2018.
[24] Loke, Seng. (2015). The Internet of Flying-Things: Opportunities and Challenges with Airborne Fog Computing and Mobile Cloud in the Clouds.
[25] T. Cheng, A. Keintola, and Blue Danube Systems, "Why Does Spectral Efficiency Matter?" [Online]. Available: https://www.bluedanube.com/blog/2018/07/23/why-does-spectral-efficiency-matter/.
[26] Liu, R., Zhang, Z., Jiao, Y., Yang, C., & Zhang, W. (2019). Study on Flight Performance of Propeller-Driven UAV. International Journal of Aerospace Engineering, 2019, 1-11. doi:10.1155/2019/6282451
[27] 'Packings of equal circles in fixed-size containers with maximum packing density" [Online] Available: http://www.packomania.com/.

.